\def\unit{\relax{\rm 1\kern-.29em l}}

\input harvmac
\input tables 
\input epsf.tex

\noblackbox



\newcount\figno
\figno=0
\def\fig#1#2#3{
\par\begingroup\parindent=0pt\leftskip=1cm\rightskip=1cm\parindent=0pt
\baselineskip=9pt
\global\advance\figno by 1
\midinsert
\epsfxsize=#3
\centerline{\epsfbox{#2}}
\vskip 12pt
{\bf Fig \the\figno} #1\par
\endinsert\endgroup\par
}
\def\figlabel#1{\xdef#1{\the\figno}}





\lref\ISS{
  K.~Intriligator, N.~Seiberg and D.~Shih,
  ``Dynamical SUSY breaking in meta-stable vacua,''
  {\tt hep-th/0602239}.
}

\lref\Kutasov{
  D.~Kutasov,
  ``A comment on duality in ${\cal N}=1$ supersymmetric non-Abelian gauge theories,''
  Phys.\ Lett.\ B {\bf 351}, 230 (1995);
  {\tt 9503086}.
}

\lref\Aharony{
  O.~Aharony, J.~Sonnenschein and S.~Yankielowicz,
  ``Flows and duality symmetries in ${\cal N}=1$ supersymmetric gauge theories,''
  Nucl.\ Phys.\ B {\bf 449}, 509 (1995);
  {\tt hep-th/9504113}.
}

\lref\KS{
  D.~Kutasov and A.~Schwimmer,
  ``On duality in supersymmetric Yang-Mills theory,''
  Phys.\ Lett.\ B {\bf 354}, 315 (1995);
  {\tt hep-th/9505004}.
}

\lref\KSS{
  D.~Kutasov, A.~Schwimmer and N.~Seiberg,
  ``Chiral rings, singularity theory and electric-magnetic duality,''
  Nucl.\ Phys.\ B {\bf 459}, 455 (1996);
  {\tt hep-th/9510222}.
}

\lref\triangle{
  M.~J.~Duncan and L.~G.~Jensen,
  ``Exact tunneling solutions in scalar field theory,''
  Phys.\ Lett.\ B {\bf 291}, 109 (1992).
}

\lref\SQCD{
  N.~Seiberg,
  ``Electric - magnetic duality in supersymmetric nonAbelian gauge theories,''
  Nucl.\ Phys.\ B {\bf 435}, 129 (1995)
  [arXiv:hep-th/9411149].
}

\lref\ElitzurHC{
  S.~Elitzur, A.~Giveon, D.~Kutasov, E.~Rabinovici and A.~Schwimmer,
  ``Brane dynamics and ${\cal N} = 1$ supersymmetric gauge theory,''
  Nucl.\ Phys.\ B {\bf 505}, 202 (1997);
  {\tt hep-th/9704104}.
}

\lref\Luca{
  L.~Mazzucato,
   ``Chiral rings, anomalies and electric-magnetic duality,''
  JHEP {\bf 0411}, 020 (2004); {\tt 
  hep-th/0408240}.
}



\vskip .2in  

\newbox\tmpbox\setbox\tmpbox\hbox{\abstractfont }
\Title{\vbox{\baselineskip12pt \hbox{CALT-68-2602}}}
{\vbox{\centerline{Landscape of Supersymmetry Breaking Vacua}  
\centerline{}
\centerline{in Geometrically Realized Gauge Theories}}}
\vskip 0.2cm

\centerline{Hirosi Ooguri  and  Yutaka Ookouchi}
\bigskip
\centerline{California Institute of Technology}
\smallskip
\centerline{Pasadena, CA 91125, USA}

\vskip 1.3cm

\noindent

We study vacuum structure of ${\cal N}=1$ supersymmetric
quiver gauge theories which can be realized geometrically by
D brane probes wrapping cycles of local Calabi-Yau three-folds.
In particular, we show that the $A_2$ quiver theory 
with gauge group $U(N_1) \times U(N_2)$ with 
${1\over 2} N_1 < N_2 < {2\over 3} N_1$ has a regime with 
an infrared free description that is partially magnetic 
and partially electric. Using this dual description, we 
show that the model has a landscape of inequivalent 
meta-stable vacua where supersymmetry is dynamically broken 
and all the moduli are stabilized. Each vacuum has distinct
unbroken gauge symmetry. The gaugino masses are generated 
by radiative corrections, and we are left with the bosonic 
pure Yang-Mills theory in the infrared. We also identify the supersymmetric 
vacua in this model using their infrared free descriptions 
and show that the decay rates of the supersymmetry breaking vacua 
can be made parametrically small.

\bigskip\bigskip
\Date{June, 2006}

\newsec{Introduction}

The discovery \ISS\ that simple supersymmetric gauge theories 
such as the ${\cal N}=1$ supersymmetric QCD with massive flavors 
have meta-stable vacua with broken supersymmetry may lead to a new 
paradigm in phenomenological model building of supersymmetric 
extensions of the Standard Model. Moreover, techniques developed 
in \ISS\ allow us to study non-perturbative aspects of these models
that are not protected by supersymmetry. 
In this paper, we will try to understand how generic such phenomena 
are by applying their method to a class of quiver gauge theories.
Not all models exhibit supersymmetry breaking vacua 
in regions that are accessible with our current technology.
It is therefore interesting to note that these phenomena
naturally happen in models that arise as low energy limits 
on D branes wrapping cycles in local Calabi-Yau manifolds.\foot{
See \ref\FrancoES{
  S.~Franco and A.~M.~Uranga,
  ``Dynamical SUSY breaking at meta-stable minima from D-branes at obstructed
  geometries,''
  {\tt hep-th/0604136}.
} for another class of string inspired models 
with meta-stable supersymmetry breaking vacua.} 
In the models studied in this paper, field content and gauge symmetry
emerging from string theory conspire to ensure that 
the supersymmetry breaking configurations are
locally stable in all directions. 

One of the models we study in this paper is the ${\cal N}=1$
supersymmetric quiver gauge theory associated to the 
$A_2$ Dynkin diagram with the gauge group $U(N_1) \times U(N_2)$.
This model can be viewed as the ${\cal N}=2$ supersymmetric
$A_2$ quiver gauge theory with the supersymmetry broken to
${\cal N}=1$ by superpotentials for the adjoint scalar fields. 
This model arises naturally from
type IIB string theory as the low energy limit of D5 brane probes 
wrapping 2-spheres in the local Calabi-Yau three-fold which 
is an $A_2$ fibration on a plane \ref\geometric{ 
  F.~Cachazo, S.~Katz and C.~Vafa, 
  ``Geometric transitions and ${\cal N}=1$ quiver theories,'' 
  {\tt hep-th/0108120}. 
}. The superpotentials encode the
geometric data on the fibration. We assume that the model is 
in the asymptotic free regime in the electric description.

\lref\Seibergone{
  N.~Seiberg,
  ``Exact results on the space of vacua of four-dimensional SUSY gauge
  theories,''
  Phys.\ Rev.\ D {\bf 49}, 6857 (1994);
  {\tt hep-th/9402044}.
}

\lref\Seibergtwo{
  N.~Seiberg,
  ``Electric-magnetic duality in supersymmetric nonAbelian gauge theories,''
  Nucl.\ Phys.\ B {\bf 435}, 129 (1995);
  {\tt hep-th/9411149}.
}

When $N_1 < N_2$, it was shown in \ref\CachazoSG{
  F.~Cachazo, B.~Fiol, K.~A.~Intriligator, S.~Katz and C.~Vafa,
  ``A geometric unification of dualities,''
  Nucl.\ Phys.\ B {\bf 628}, 3 (2002);
  {\tt hep-th/0110028}.
} that this model has a Seiberg-like dual \refs{\Seibergone, \Seibergtwo}
which is 
the same $A_2$ quiver theory but with the different gauge group 
$U(N_2-N_1) \times U(N_2)$. 
In the type IIB string language, 
this duality is the Weyl reflection symmetry 
of the $A_2$ Dynkin diagram. This duality also has the following 
field theoretical interpretation. Let us assume that the strong 
coupling scales $\Lambda_1$ and $\Lambda_2$ for the $U(N_1)$ and $U(N_2)$ 
obey $\Lambda_1 \gg \Lambda_2$. (Geometrically, 
this means that we choose one of the 2-spheres to be much 
smaller than the other.) We can then performs the Kutasov-type 
duality \refs{\Kutasov, \Aharony, \KS, \ElitzurHC, \Luca} 
on the $U(N_1)$ to land 
on a theory with the gauge group 
$U\left((k-1)N_2 - N_1\right) \times U(N_2)$, where $k$ is the highest power
in the superpotential $W_1(X_1)$ for the adjoint scalar $X_1$
for the electric gauge group $U(N_1)$.
The $F$-term constraints force the gauge 
symmetry to be spontaneously broken to 
$U(N_2 - N_1) \times U(N_2)$, which is the dual gauge group 
identified in \CachazoSG .  

In this paper, we will consider the same model but in a different
regime. When the superpotential $W_1(X_1)$ is cubic, the
Kutasov-type dual becomes infrared free for both gauge groups when 
${1\over 2} N_1 < N_2 < {2\over 3} N_1$.\foot{For a more general
potential $W_1(X_1) \sim X_1^{k} + \cdots$, the corresponding
condition is ${1\over k-1}N_1 < N_2 < {2 \over 2k-3} N_1$.
However, this condition is not compatible with the asymptotic freedom of
the electric description unless
$k \leq 3$. We restrict our attention to the case when
$W_1(X_1)$ is cubic so that the
original electrical description is ultraviolet complete.} 
We find that, in this case, the 
$D$ and $F$-term constraints have no solutions near the origin 
of the field space. On the other hand, there are local 
minima of the classical $D$ and $F$-term potentials. Each
of these minima has massless chiral multiplets at the tree level, 
but they all become massive by the one-loop effective potential.
Thus, there are no flat directions at these meta-stable vacua.   
Higher order corrections to the potential can be 
made parametrically small because of the infrared freedom.

Each of the isolated meta-stable vacua has distinct unbroken 
gauge symmetry of the form $U(r_1) \times U(r_2) 
\times U(N_1-N_2)$ with $r_1+r_2=2N_2-N_1$. At the 
supersymmetry breaking scale, all chiral multiplets become massive.
Unless both $N_1$ and $N_2$ are even, the chiral anomaly combined
with the superpotential breaks the R symmetry completely. 
Thus, the supersymmetry breaking also generates gaugino
massives by radiative corrections, and we are left with 
the bosonic Yang-Mills theory for the unbroken gauge symmetry. 
In particular, the broken 
supersymmetry is not restored in the infrared. This gives the landscape 
of supersymmetry breaking vacua, each of which is characterized
by the gauge symmetry breaking pattern.

We find that there are supersymmetric vacua away from
the origin of the field space in the dual description.
This is consistent with the electric description of the
model and establishes the connection of the two descriptions. 
The identification of the supersymmetric vacua in
the dual description also allows us to estimate the decay rates 
of the supersymmetry breaking vacua into the supersymmetric
vacua. We find that the decay rates can be made parametrically
small. Since the supersymmetry breaking vacua all have the same energy, 
the transition probabilities among them are equal to zero.

We also study the $A_2$ quiver theory with $SU(N_1) \times SU(N_2)$
gauge group, which has an analogous infrared free dual description. 
In the $U(N_1) \times U(N_2)$ model, the energies of 
the supersymmetry breaking vacua are degenerate. We find that
the degeneracy is lifted in the $SU(N_1) \times SU(N_2)$ model. 

Generalization of our results to gauge theories associated to
more general quiver diagrams is currently under investigation \ref\inproress{
H.~Ooguri and Y.~Ookouchi, in progress.}. It would be interesting
to find out whether the existence of meta-stable vacua with
broken supersymmetry is a generic phenomenon for this class
of gauge theories realized as low energy limits of string theory.

Study of meta-stable vacua with broken supersymmetry may
provide a new insight into the flux compactification of 
string theory of the type pioneered in 
\lref\GiddingsYU{
  S.~B.~Giddings, S.~Kachru and J.~Polchinski,
  ``Hierarchies from fluxes in string compactifications,''
  Phys.\ Rev.\ D {\bf 66}, 106006 (2002);
  {\tt hep-th/0105097}.
}
\lref\KachruAW{
  S.~Kachru, R.~Kallosh, A.~Linde and S.~P.~Trivedi,
  ``De Sitter vacua in string theory,''
  Phys.\ Rev.\ D {\bf 68}, 046005 (2003);
  {\tt hep-th/0301240}.
}
\refs{\GiddingsYU, \KachruAW}.
In this connection, it would also be interesting to
study properties of the meta-stable supersymmetry 
breaking vacua of the gauge theories from the geometric
point of view
of string theory. This has been attempted earlier for 
models with stable supersymmetry breaking vacua, 
for example in 
\ref\deBoerBY{
  J.~de Boer, K.~Hori, H.~Ooguri and Y.~Oz,
  ``Branes and dynamical supersymmetry breaking,''
  Nucl.\ Phys.\ B {\bf 522}, 20 (1998);
  {\tt hep-th/9801060}.
} using the M theory fivebrane description 
of ${\cal N}=1$ supersymmetric gauge theories 
\lref\mfiveone{
  K.~Hori, H.~Ooguri and Y.~Oz,
  ``Strong coupling dynamics of four-dimensional ${\cal N} = 1$ 
gauge theories from  M
  theory fivebrane,''
  Adv.\ Theor.\ Math.\ Phys.\  {\bf 1}, 1 (1998);
  {\tt hep-th/9706082}.
}
\lref\mfivetwo{
E.~Witten,
  ``Branes and the dynamics of QCD,''
  Nucl.\ Phys.\ B {\bf 507}, 658 (1997);
  {\tt hep-th/9706109}.
}
\lref\mfivethree{
  A.~Brandhuber, N.~Itzhaki, V.~Kaplunovsky, J.~Sonnenschein and S.~Yankielowicz,
  ``Comments on the M theory approach to ${\cal N} = 1$ SQCD and brane dynamics,''
  Phys.\ Lett.\ B {\bf 410}, 27 (1997);
  {\tt hep-th/9706127}.
}
\refs{\mfiveone,\mfivetwo,\mfivethree}. It would be
worth revisiting this issue from the new perspective
that is emerging. 

\newsec{$U(N_1) \times U(N_2)$ Gauge Theories}

In this section, we consider the $U(N_1) \times U(N_2)$
quiver gauge theory as described in the introduction. The
chiral multiplets of the theory are $X_1$ and $X_2$ which
are adjoint in $U(N_1)$ and $U(N_2)$ respectively and 
$Q_{12}$ and $Q_{21}$ which are 
bi-fundamental in $U(N_1) \times U(N_2)$. 
In the ${\cal N}=1$ language, the total superpotential 
is given by
\eqn\totalpotential{ W = W_1(X_1) + W_2(X_2) + \tr \ Q_{21} X_1 Q_{12} +
 \tr \ Q_{12} X_2 Q_{21}. }
The last two terms are inherited from the ${\cal N}=2$
quiver theory. The superpotentials $W_1(X_1)$ and $W_2(X_2)$
reduce the supersymmetry to ${\cal N}=1$.
We assume that $2N_1 - N_2 >0$ and $2N_2 - N_1  >0$ so that 
the model is asymptotically free. The strong coupling 
scales for $U(N_1)$ and $U(N_2)$ are denoted
by $\Lambda_1$ and $\Lambda_2$ respectively. 

\subsec{Magnetic dual}

Suppose $\Lambda_1 \gg \Lambda_2$. When $N_1 < N_2$, it is well-known that
the model has a magnetic dual with the gauge group 
$U(N_2-N_1) \times U(N_2)$.  Geometrically, this duality is 
the Weyl reflection symmetry of the $A_2$ Dynkin diagram 
relating inequivalent blow-ups of the $A_2$ singularity 
\CachazoSG.\foot{This is closely related to the duality cascade of
 \ref\KlebanovHB{
  I.~R.~Klebanov and M.~J.~Strassler,
  ``Supergravity and a confining gauge theory: Duality cascades and
  $\chi$SB-resolution of naked singularities,''
  JHEP {\bf 0008}, 052 (2000);
  {\tt hep-th/0007191}.
}, which corresponds to the affine $\widehat{A}_1$ case.} 

Here, we will consider another region $N_2 < {2 \over 3} N_1$.
We choose the superpotential terms to be
\eqn\specialpotential{ W_1(X) = t_0\  \tr \left( {1\over 3} X^3 - t_1^2 X
\right), ~~W_2(X) = 0,}
where $t_0$ and $t_1$ are constant parameters. As we mentioned in
Footnote 2 in the introduction, we choose the highest power 
in $W_1(X_1)$ to be cubic. For simplicity, we set $W_2(X_2)=0$, but it
is straightforward to consider the case when $W_2$ is a general
polynomial since we assume the $U(N_2)$ gauge sector is weakly
coupled.\foot{In particular, the matter content of the dual theory
is independent of $W_2(X_2)$ but depends on $W_1(X_1)$.} A quadratic term
is not included in $W_1(X_1)$ since it can be removed by a combined 
shift of $X_1 \rightarrow X_1 + c\ \unit$ and $X_2 \rightarrow 
X_2 - c\ \unit$ for some constant $c$. (We need to shift $X_2$
simultaneously  
in order to keep the last two terms in \totalpotential\ unchanged.) 

To identify the dual description, we first look at the theory at 
the scale $E$ where $\Lambda_2 \ll E$.
There, the $U(N_2)$ gauge 
sector of theory is weakly coupled and can be treated
as a spectator. 
On the other hand, the $U(N_1)$ gauge sector can be strongly coupled.
This sector consists of the $U(N_1)$ gauge field coupled to
$N_2$ fundamental fields $(Q_{12}, Q_{21})$ and  
the adjoint field $X_1$ with the superpotential $W_1(X_1)$.
The weakly coupled magnetic dual of this model has been 
identified in \refs{\Kutasov, \Aharony, \KS, \ElitzurHC} and consists  
of the gauge field for $U(2N_2 - N_1)$,
$N_2$ fundamental fields $(q_{12}, q_{21})$, one adjoint 
field $Y$, and two neutral fields $M, M'$ which are 
in the adjoint representation of $U(N_2)$. The superpotential
for the dual theory is 
\eqn\newpotential{  \eqalign{
\widetilde{W} = 
-t_0\ \tr\left({1\over 3} Y^3 -t_1^2 Y \right) 
+  \ {t_0 \over \mu_0^2} \ & \tr \left(M q_{21} q_{12}
+ M' q_{21} Y q_{12}  \right)  \cr
  + & \tr \left(M + X_2 M' \right),}}
where $\mu_0$ relates the strong coupling
scale $\Lambda_1$ of the electric gauge group $U(N_1)$
to the scale $\widetilde{\Lambda}_1$ of	its magnetic dual
$U(2N_2-N_1)$ as
$$ \Lambda_1^{2N_1-N_2} \widetilde{\Lambda}_1^{3N_2-2N_1} 
= \left({\mu_0 \over t_0}\right)^{2N_2}. $$

Since the dual theory has one adjoint field $Y$ and since 
$(q_{12}, q_{21})$ can be regarded as $N_2$ fundamental
fields with respect to the $U(2N_2-N_1)$, the coefficient
of the beta-function is given by 
$$ b_{U(2N_2-N_1)}= 3(2N_2 - N_1) - N_2 - (2N_2-N_1) = 3N_2 - 2 N_1, $$
and it becomes negative when $N_2 < {2\over 3} N_1$. 
At the same time, since there are three adjoint fields $M, M', X_2$
and since $(q_{12}, q_{21})$ count as $(2N_2-N_1)$ fundamental 
fields for $U(N_2)$, we have 
$$ b_{U(N_2)} = 3 N_2 - (2N_2 - N_1) - 3 N_2 = N_1- 2N_2. $$
Thus, both $U(2N_2-N_1)$ and $U(N_2)$ gauge couplings are infrared free when
$$  {1\over 2} N_1 < N_2 < {2\over 3} N_1 .$$ 
Moreover, in the magnetic dual, 
we land on the weak coupling regime of the $U(N_2)$ gauge group. To
see this, we note that the scale $\tilde\Lambda_2$ of this gauge
group is related to its original scale $\Lambda_2$ in the electric
description by the matching condition,
$$ \left({\Lambda_2 \over E}\right)^{2N_2-N_1} =
 \left({\tilde{\Lambda}_2 \over E}\right)^{-(2N_2-N_1)}, $$
namely,
$$ E = \sqrt{\Lambda_2 \tilde{\Lambda}_2}. $$
Thus, when we start with the weak coupling regime $E \gg \Lambda_2$
in the asymptotic free electric description, we land on the
the weak coupling regime $E \ll \tilde{\Lambda}_2$, well below the
Landau pole in the infrared free magnetic description. 

One may consider generalizing this construction by allowing
higher powers of $X_1$ in the superpotential as 
$$ W_1(X_1) = \ t_0  \ \tr X_1^k + \cdots. $$
The Kutasov-type dual has the gauge group $U((k-1)N_2 - N_1) \times
U(N_2)$, and the coefficients of the beta functions are
$$ \eqalign{ b_{U((k-1)N_2 - N_1)}
 & = (2k-3) N_2 - 2N_1 ,\cr 
b_{U(N_2)} & =N_1 - (k-1)N_2 .\cr}$$
If we require that the electric description is 
asymptotic free, in particular $2N_2 > N_1$,
we have $b_{U((k-1)N_2-N_1)} > 0$ for $k > 3$. 
Thus, the case with $k=3$ is special in the sense
that the electric description is ultraviolet
complete and the magnetic description is infrared free. 

Since the dual theory is in the infrared free range
for $k=3$, the K\"ahler potential 
is regular around the origin of the field space and can be expanded as
$$
K=\left({1 \over \alpha \Lambda_1^2}\right)^2 
M^{\dagger}M + \left({1\over \alpha' \Lambda_1}\right)^2
M'^{\dagger}M'+ {1\over \beta^2} (q^{\dagger}q+\tilde{q}^{\dagger}\tilde{q})
+ {1\over \gamma^2}  Y^{\dagger}Y
+{1\over \gamma'^2} X_2^{\dagger}X_2,$$
for some unknown coefficients $\alpha, \alpha', \beta, \gamma, \gamma'$.
It is reasonable to 
assume that these coefficients are regular for some range of the parameters,
$t_0$, $t_1$, $\mu_0$, $etc$. 
Trace symbols are implicit in the above. 
Since $M$ and $M'$ are identified with 
$Q_{12}X_1 Q_{21}$ and $Q_{21}Q_{12}$ in the
electric description, their dimensions are not equal to $1$, which is why
their kinetic terms are divided by the electric scale $\Lambda_1$. 
We will find it useful to rescale the fields, $e.g.$ as 
$M \rightarrow \alpha \Lambda_1^2 M$, so that their coefficients
in the K\"ahler potential are normalized to be $1$.  
The superpotential \newpotential\ after this rescaling becomes 
\eqn\newpotentialII{ 
\eqalign{ \widetilde{W} = 
-t_0 \ \tr\left({1 \over 3} \gamma^3 Y^3 - t_1^2 \gamma Y \right) 
&+ \ h \ \tr \left[ M\left( q_{21} q_{12} - \mu^2 \unit\right)\right]
\cr &+ \ h' \ \tr \left[M' \left( q_{21} \gamma Y q_{12} - \mu'^2 X_2 \right)\right]
,\cr}}
where
$${h= \alpha \beta^2\ t_0 \ { 
\Lambda_1^2 \over \mu_0^2}, \quad
h'=\alpha' \beta^2\ t_0 \ {\Lambda_1 \over \mu_0^2}, \quad  
\mu^2=-{ 1 \over \beta^2}\ {\mu_0^2\over t_0} , \quad  
\mu'^2=-{\gamma' \over \beta^2}
\ {\mu_0^2\over t_0}
}$$
In the following, we will analyze the vacuum structure of 
the gauge theory using this dual description.

It is instructive to compare the above construction with that for
$N_2 > N_1$. In this case, 
the dual $U(2N_2 - N_1)$ gauge sector is asymptotically free. 
Moreover, since $2N_2 - N_1 > N_2$, the $F$-term
constraints require partial Higgsing of the gauge symmetry,
and we end up with a magnetic
dual with the ${\cal N}=1$ quiver theory with the gauge group 
$U(N_2 - N_1) \times U(N_2)$, resulting in the cascade structure as
shown in \CachazoSG. In contrast, in our case when 
$N_2 < {2 \over 3} N_1$, the rank condition discussed below 
makes it impossible to solve the $F$-term constraints, leading
to supersymmetry breaking.

\subsec{Tree-level potential}

The $F$-term conditions for the dual theory are
\eqn\Ftermeq{\eqalign{
q_{21}q_{12}-\mu^2\ \unit_{N_2} &=0,\cr
q_{21}\gamma Y q_{12}-\mu'^2 X_2&=0,\cr
q_{12}M=Mq_{21}&=0,\cr 
\gamma^2 Y^2  - t_1^2\ \unit_{2N_2-N_1} &=0, \cr
M' & = 0.}}
Note that $q_{12}$ is a matrix of size $(2N_2 - N_1) \times N_2$.
Since $2N_2 - N_1 < N_2$, the rank of $q_{12}$ is
$(2N_1 - N_2)$ at most, and the first equation can
never be satisfied. This is the rank condition mechanism of \ISS. 
Thus, the supersymmetry is broken at the tree level
in this description. What we need to check is whether
there are field configurations that are locally stable. 

The $F$-term potential contains a term 
proportional to $\tr \ \big| 
q_{21} q_{12} - \mu^2 \unit_{N_2} \big|^2 $.
This can be minimized by setting 
\eqn\minq{q_{21}= 
\pmatrix{\varphi_0 \cr 0},\quad  
q_{12}=\pmatrix{\widetilde{\varphi}_0 &0 }}
where $\varphi_0$ and $\tilde \varphi_0$ are $(2N_2-N_1)\times (2N_2-N_1)$ 
matrices satisfying
\eqn\whatvarphi{
 \widetilde{\varphi}_0\varphi_0=\mu^2 \ \unit_{2N_2-N_1}.}
For this configuration, 
$$ 
\tr \ \left| {\partial W \over \partial M}\right|^2
= |h|^2 \ \tr \ \left| 
q_{21} q_{12} - \mu^2 \unit_{N_2} \right|^2 
= (N_1-N_2)|h\mu^2|^2 . $$
As we will see below, the remaining $F$-term conditions and
all the $D$-term conditions can be solved for this
choice of $(q_{12}, q_{21})$. Moreover, 
the minimum value of $\tr \ |\partial W/\partial M|^2$ 
depends only on the parameters
of the model and not on the field variables, and cannot
be minimized further.
Thus, this gives the minimum value of the tree-level potential,
\eqn\treepot{
 V_{{\rm min}}^{{\rm (tree)}} = (N_1-N_2)|h\mu^2|^2 = 
(N_1-N_2) |\alpha \Lambda_1^2|^2.}
For this choice of $(q_{12}, q_{21})$, the rest
of the $F$-term conditions in
\Ftermeq\ can be solved by setting
\eqn\Vmin{M =\pmatrix{0&0 \cr 0& \Phi_0}, \quad
 M'=0, \quad 
X_2={1 \over \mu'^2}\,\widetilde{\varphi}_0 \gamma Y \varphi_0,
\quad \gamma^2 Y^2 =   t_1^2 \ \unit_{2N_2-N_1}
 }
where $\Phi_0$ is an arbitrary $(N_1-N_2)\times(N_1-N_2)$ matrix.

Let us turn to the $D$-term conditions. It is well-known that,
for a supersymmetric theory with gauge group $G$, any field configuration
solving the $F$-term conditions can be mapped by a complexified gauge
transformation $G_c$ to a unique solution to the $D$-term constraints
modulo the $G$ gauge transformation 
\ref\LutySD{
  M.~A.~Luty and W.~I.~Taylor,
  ``Varieties of vacua in classical supersymmetric gauge theories,''
  Phys.\ Rev.\ D {\bf 53}, 3399 (1996);
  {\tt hep-th/9506098}.
}. In our case, since $Y$ satisfies the last equation in \Vmin,
we can use $GL(2N_2-N_1,C)$ transformation to diagonalize it in the form,
\eqn\ydiag{ \gamma Y = {\rm diag}(t_1, \cdots, t_1, -t_1, \cdots, -t_1). }
This configuration breaks $U(2N_2-N_1)$ into $U(r_1) \times U(r_2)$
where $r_1+r_2=2N_2-N_1$ and $r_2$ is the number of minus signs in the above.
We can also use $GL(2N_2 - N_1, C) \subset GL(N_2, C)$ to 
transform a solution to \whatvarphi\ 
into
\eqn\phidiag{ \varphi_0 = \tilde\varphi_0 = \mu\ \unit_{2N_2-N_1}.}
Finally, $\Phi_0$ in \Vmin\ can be diagonalized by the remaining
gauge symmetry $GL(N_1-N_2,C) \subset GL(N_2,C)$.
It is straightforward to verify that the resulting field configuration
solves the $D$-term constraints. 

To summarize, we found that the tree-level $D$ and $F$-term potential
is minimized by \minq\ and \Vmin\ where $Y$ and $\varphi_0, \widetilde\varphi_0$
are fixed as in \ydiag\ and \phidiag\ and $\Phi_0$ is also diagonalized.
The only flat directions that are not fixed by the tree-level
potential are the eigenvalues of $\Phi_0$. In the following, we will
compute a one-loop effective potential for $\Phi_0$ to show that
all its eigenvalues are stabilized at one-loop. 

\subsec{One-loop effective potential}

We are going to compute the one-loop effective potential for $\Phi_0$, 
which is the only massless chiral multiplet at the tree-level. 
Let us parametrize the fluctuations around one of the tree-level vacua as, 
\eqn\fluctuation{\eqalign{q_{12}& =\mu \pmatrix{{\unit } & 0}
+ \pmatrix{\sigma_1 &  \phi_1 } , \quad
q_{21}=\mu \pmatrix{ {\unit } \cr 0 }
+ \pmatrix{ \sigma_2 \cr  \phi_2 } \cr 
M&=\pmatrix{ 0 & 0 \cr 0 & \Phi_0 } + 
\pmatrix{ \sigma_3 &  \phi_4 \cr  \phi_3 & 
 \sigma_4 }, \quad M'=\pmatrix{ \sigma_5 &  \phi_6 \cr  \phi_5 & \sigma_6 },
\cr
X_2& ={\mu^2 \over \mu'^2} \pmatrix{\gamma  Y_0   & 0
\cr 0 & 0 } + \pmatrix{
  \sigma_7 &  \phi_8 \cr \phi_7 &  \sigma_8 }, \quad
Y= Y_0  + \sigma_9,
}}
and expand the action to the quadratic order in $\phi$'s and $\sigma$'s.
Here $Y_0$ is the vacuum value of $Y$ given by \ydiag . 
We then perform the Gaussian integral for $\phi$'s 
and $\sigma$'s to compute the one-loop effective potential
for $\Phi_0$. Since the vacuum configuration satisfies 
the $D$-term condition, one-loop contributions from
the vector multiplets are canceled, even though some of their
masses depend on $\Phi_0$. Since one-loop diagrams are planar, 
the effective potential $V_{{\rm eff}}^{(1)}(\Phi_0)$ for $\Phi_0$ 
should be expressed in terms of a single trace. The quadratic 
term in the expansion of $V_{{\rm eff}}^{(1)}$ around $\Phi_0=0$ should 
then be of the form $ V_{{\rm eff}}^{(1)} \sim \tr \ \Phi_0^\dagger \Phi_0$
by the $U(N_2)$ symmetry. This means that all the eigenvalues of $\Phi_0$
get the same mass at one-loop. Without a loss of generality, we can
set
$$ \Phi_0 = X  \ \unit_{N_1 - N_2}.$$
We can then expand 
$V_{{\rm eff}}^{(1)}(X)$ in powers of $X$ and look
at the quadratic term to find the mass for $\Phi_0$.  

The one-loop effective potential for $X$ 
is then given by
\eqn\oneloopeff{
V_{\rm eff}^{(1)}(X)={1\over 64 \pi^2}S\Tr {\cal M}^4 
\log {{\cal M}^2 \over \Lambda^2}
\equiv {1\over 64 \pi^2}\sum \left(  m_B^4 \log {m^2_B \over \Lambda^2}
-  m_F^4 \log {m^2_F \over \Lambda^2} \right),
}
where $\Lambda$ is the ultraviolet cutoff parameter, and
$m_B$ and $m_F$ are the masses for the bosons and fermions 
that are given by expanding the $D$ and $F$-term potentials in 
the quadratic order in $\phi$'s and $\sigma$'s. The effective
potential is a function of $X$ because the masses depend on it. 
The $D$-term potential 
for the $U(2N_2-N_1)$ gauge symmetry does not contain fields that are
directly coupled to the supersymmetry breaking sector, and therefore
it does not contribute to the effective potential. 
The $D$-term potential for the $U(N_2)$ gauge symmetry contains
such fields, but their effects are suppressed since the gauge coupling
for the $U(N_2)$ is weak in this energy scale. 
Thus, we only consider the $F$-term potential to compute the one-loop
effective potential. 

The mass matrices we use to 
evaluate \oneloopeff\ are therefore given in terms of the superpotential $W$
as
\eqn\massmatrix{
m_B^2=\pmatrix{ W^{\dagger ac}W_{cb} & W^{\dagger abc}W_{c} 
\cr W_{abc}W^{\dagger c} & W_{ac}W^{\dagger cb} },
\qquad m_{F}^2=\pmatrix{ W^{\dagger ac}W_{cb} & 0 \cr 0 & W_{ac}W^{\dagger cb} },
}
where $a,b,c,..$ represent the fluctuations $\phi$'s and $\sigma$'s
in \fluctuation ,  and the derivatives of $W$ are evaluated at the tree-level 
vacuum configuration. If $W_{abc}W^{\dagger c}= 0$, the supertrace
in \oneloopeff\ vanishes by cancellation between bosons and fermions
contributions. Since the only field $c$ with 
$W_c\neq 0$ is $c=\sigma_4$ and since $\sigma$'s do not have
non-zero $W_{abc}$ with $\sigma_4$, only the $\phi$ fluctuations
contribute to the one-loop effective potential. 

Each $\phi$ in the above is an $(2N_2-N_1) \times (N_1-N_2)$
matrix. From this, it follows that the classical action for 
the quadratic fluctuations is a sum of $(2N_2-N_1)(N_1-N_2)$ 
copies of the O'Raifeartigh-type model with the superpotential 
\eqn\ORai{\eqalign{
W=hX\left(\phi_1 \phi_2-\mu^2\right)& +h \mu\  (\phi_1 \phi_3 + 
\phi_2\phi_4)\cr 
&+h' \mu \gamma Y (\phi_1  \phi_6 + \phi_2 \phi_5)
-h' \mu'^2 ( \phi_6 \phi_7 + \phi_5 \phi_8).
}
}
Here $\gamma Y = + t_1$ for $r_1 (N_1-N_2)$ of them,
and $\gamma Y = -t_1$ for the rest. 

We can now evaluate the one-loop effective potential \oneloopeff\
and expand it in powers of $X$. It turns out that the potential
is independent of $(r_1, r_2)$. Thus, both the vacuum energy
and the mass for $X$ are the same for all the vacua. 
The constant term in $V_{\rm eff}^{(1)}$ gives the one-loop correction to the
vacuum energy. Combining it the tree level result \treepot, 
we find that the vacuum energy is given by
\eqn\vacenergy{{\eqalign{ 
V_{\rm vac}=&(N_1-N_2)|h^2\mu^4| \cr
& + (N_1-N_2)(2N_2-N_1)\ {|h^4 \mu^4|\over 32\pi^2}\
\Big[2\log \left(|h^2 \mu^2|\Lambda^{-2}\right) 
+(2u+2v )^2 \log (2u+2v) \cr &  
 ~~~~~~~~~ +F_+({u},\, {v })^2 \log F_+({u},\, {v})  
 +F_-({u},{v })^2 \log F_-({u},\, {v }) \cr
&~~~~~~~~~   -2 F_+({2u}\,\, {2v })^2 \log F_+({2u},\, {2v }) 
-2F_-({2u},\, {2v })^2 \log F_-({2u},\, {2v})  \Big],
}}} 
where we set 
\eqn\whatxy{ \eqalign{ u&=\left| {h'^2 t_1^2 \over 2h^2}\right| 
= {1\over 2} {\alpha'^2 \over \alpha^2}
\left|{t_1^2 \over \Lambda_1^2}\right|, \cr
v & = \left|{h'^2 \mu'^4 \over 2h^2 \mu^2}\right|
= {1\over 2} {\alpha'^2 \gamma'^2 \over  \alpha^2 \beta^2} 
\left|{\mu_0^2 \over t_0 \Lambda_1^2}\right|, }}
and the function $F_\pm$ of $(u,v)$ are given by
$$F_{\pm}(u,\, v)=1+u +v \pm \sqrt{(1+u+v)^2-4v}.$$
We note that the one-loop correction to the vacuum energy is independent
of the gauge symmetry breaking pattern parametrized by $(r_1, r_2)$. 

The mass squared, $m_X^2$, can be expressed analytically as
$$ m_X^2 = (N_1-N_2)(2N_2-N_1)\ {|h^4 \mu^2|\over 16\pi^2} 
\ G(u,v), $$
for some function $G(u, v)$, where $(u,v)$ are defined in \whatxy. 
In particular, the $\log \Lambda$ terms in \oneloopeff\
are canceled out in $m_X^2$. The expression for $G$ is 
too lengthy to reproduce here. Its behavior for  $0 \leq u, v \leq 1$
is displayed in Figures 1 and 2. 
Clearly in this range the function $G(u,v)$ stays positive,
and the $X$ direction is stabilized at one-loop.
We checked numerically that $G(u,v)>0$ for a much larger
range of $u$ and $v$. In the limit $u \rightarrow \infty$, 
$G(u, v) \rightarrow 0$. Thus, the one-loop effective potential 
for $X$ becomes asymptotically shallow for large $t_1$. 
  
We observe that the dependence of $m_X^2$ on $v$ is relatively
mild. In particular, its behavior for 
$u \rightarrow 0$ is independent of $v$ as
$$G(u, v) = 4 (\log 4 - 1) + 16 (\log 2 - 1)u
+{\cal O}(u^2). $$
We can also see this graphically in Figure 2.  
The leading behavior of $m_X^2$
for $u\rightarrow 0$ coincides with that of the supersymmetric QCD
with flavors evaluated in \ISS .

\bigskip 
\centerline{\epsfxsize 3.5truein\epsfbox{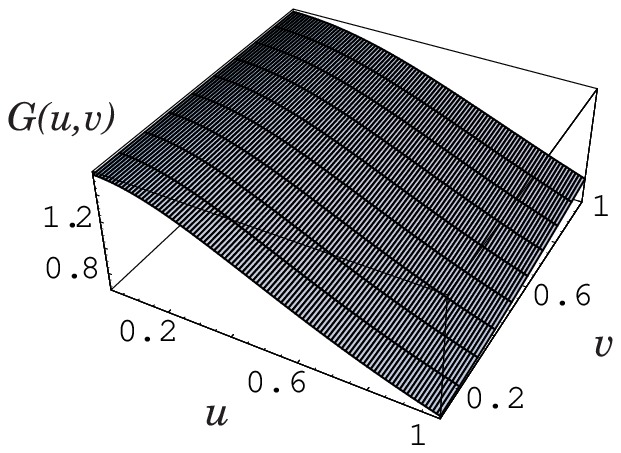}} \leftskip 2pc 
\rightskip 2pc \noindent{\ninepoint\sl \baselineskip=2pt {\bf
Fig.1} {{$G(u,v)$ for $0 \leq u, v \leq 1$}}} 
\bigskip 
\bigskip 
\centerline{\epsfxsize 3.5truein\epsfbox{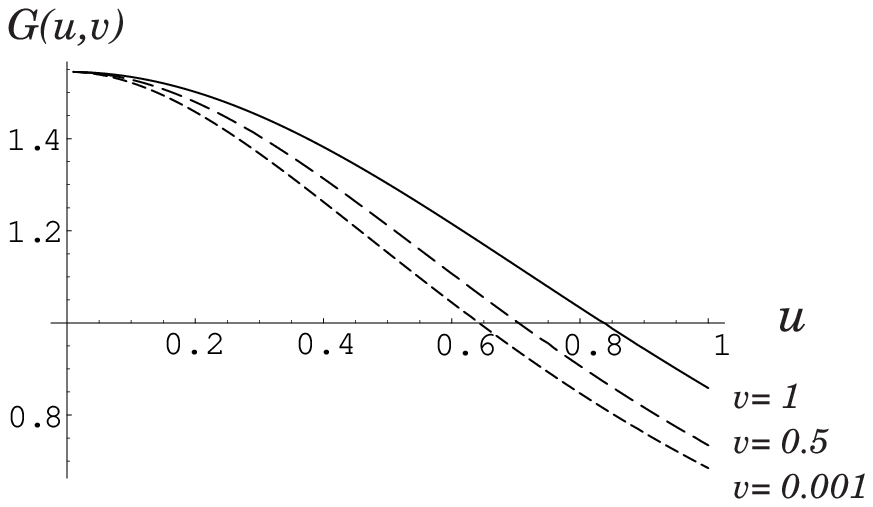}} \leftskip 2pc 
\rightskip 2pc \noindent{\ninepoint\sl \baselineskip=2pt {\bf
Fig.2} {{$G(u,v)$ shown as a function of $u$ for
$v=0.001, 0.5$ and $1$.}}} 

\leftskip 0pc 
\rightskip 0pc
\bigskip 

To conclude, we have shown that the remaining flat directions
parametrized by the eigenvalues of $\Phi_0$ are all lifted 
by the one-loop effective potential.  
It is worth mentioning that not all supersymmetric gauge theories
have meta-stable supersymmetry breaking vacua of this type.
Suppose, for example, that the $U(N_2)$ symmetry were not gauged. 
Then, we would not have the $D$-term constraint for this gauge 
symmetry, which is needed to make some of the fields massive
at the tree level. One can see that, in such a model, extra
flat directions emerge at the origin of the $\Phi_0$ space,
and they may cause the runaway behavior. A similar problem happens
to the stability of the vacua if we do not have the adjoint field
$X_2$ for this group in the electric description even if the $U(N_2)$
is gauged. It is interesting
that the geometric construction by string theory produces
exactly the right combination of field content and gauge symmetry
so that the supersymmetry breaking configurations are
locally stable in all directions. 

\subsec{Gaugino masses and the low energy limit}

We found that the chiral multiplet $\Phi_0$ gains
mass $m_X$ at the one-loop level. Thus, all the moduli
around the meta-stable vacua are stabilized. 
However, this is not the end of the story; there is unbroken 
gauge symmetry $G=U(r_1) \times U(r_2) \times U(N_1-N_2)$
with $r_1+r_2 = 2N_2-N_1$
at each of the vacua. We then need to find out the fate of the
vector multiplet for the unbroken gauge symmetry. 

Because of the term $h\ \tr M (q_{21}q_{12}
-\mu^2 \unit)$ in the superpotential $\widetilde{W}$, 
the $F$-term for the superfield $M$ is non-vanishing 
at these vacua as
$$ F_M= {\partial W \over \partial M} = h \left(q_{21}q_{12}-\mu^2 \ \unit_{N_2}\right)
 = -h \mu^2\pmatrix{ 0 & 0 \cr 0 &  ~\unit_{N_1-N_2}}, $$
where we used the vacuum values of $q_{12}$ and $q_{21}$ given 
in \minq\ and \whatvarphi . Thus, this superpotential term gives
rise to the so-called soft-breaking B-term for the bosonic components
of $(q_{12}, q_{21})$ of the form, 
\eqn\bterm{ {\cal L}_{{\rm B-term}} = h \mu^2 \ \tr
\left[ \pmatrix{ 0 & 0 \cr 0 &  ~\unit_{N_1-N_2}} q_{21}q_{12}\right] + {\rm c.c.},  }
where $q_{12}, q_{21}$ refer to the bosonic components of
the superfield. This is a part of the off-diagonal element 
in the mass matrix $m_B^2$ in \massmatrix . Moreover, 
unless both $N_1$ and $N_2$ are even, the R symmetry becomes 
trivial.\foot{To see this, we note that the superpotential given by
\totalpotential\ and \specialpotential\ breaks the $U(1)$ R symmetry
to $Z_4$. On the other hand, the chiral anomaly breaks it to 
$Z_{2n}$ where $n$ is the largest common divisor of 
$(2N_1-N2)$ and $(2N_2-N1)$. In particular, $n$ is odd unless both
$N_1$ and $N_2$ are even. In this case, the intersection of  
$Z_4$ and $Z_{2n}$ is $Z_2$, which is the fermion number
partity.} Thus, we expec that 
gaugino masses are generated by radiative corrections
\lref\Dineone{
  M.~Dine and A.~E.~Nelson,
  ``Dynamical supersymmetry breaking at low-energies,''
  Phys.\ Rev.\ D {\bf 48}, 1277 (1993);
  {\tt hep-ph/9303230}.
}
\lref\Dinetwo{
  M.~Dine, A.~E.~Nelson and Y.~Shirman,
  ``Low-energy dynamical supersymmetry breaking simplified,''
  Phys.\ Rev.\ D {\bf 51}, 1362 (1995);
  {\tt hep-ph/9408384}.
}
\lref\Dinethree{
  M.~Dine, A.~E.~Nelson, Y.~Nir and Y.~Shirman,
  ``New tools for low-energy dynamical supersymmetry breaking,''
  Phys.\ Rev.\ D {\bf 53}, 2658 (1996);
  {\tt hep-ph/9507378}.
}
\refs{\Dineone,\Dinetwo,\Dinethree}
 for the entire gauge group
$G$ except for the diagonal $U(1)$, 
for which $q_{12}, q_{21}$ are neutral. 

The massive gauginos decouple in the infrared, and 
we are then left with the bosonic pure Yang-Mills theory
for the gauge group $G' = G / U(1)$ and
the free $U(1)$ supersymmetric Yang-Mills theory. 
Since the original electrical description of the theory
contains no field charged with respect to the 
diagonal $U(1)$, this abelian Yang-Mills theory is decoupled 
from the beginning of the story and we can ignore it. 
Thus, the low energy limit is the bosonic pure Yang-Mills
theory for gauge group $G'$, and the supersymmetry is 
broken. The strong coupling scale of the bosonic Yang-Mills theory 
is determined by the matching condition at the supersymmetry breaking scale.
This sets the lowest energy scale of the model, and  
the lowest excitations of the model around the meta-stable 
vacua are glueballs of the Yang-Mills theory.

\subsec{Supersymmetric vacua and decay rates of meta-stable vacua}

In addition to the meta-stable vacua with broken supersymmetry that we
have looked at so far, the model also has supersymmetric vacua
as one can see in the original electric description 
\geometric. 
Here, we will show that these supersymmetric vacua 
can also be found in the dual description
within the region of its validity, establishing the connection of 
the two description. We will use this result to estimate the
decay rates of the supersymmetry breaking vacua into the supersymmetric 
vacua.

Following \ISS, we look for the supersymmetric vacua where $M$ is large. 
There $(q_{12}, q_{21})$ are heavy and can be integrated out. 
Thus, we are left with the superpotential,
\eqn\magneticsup{
\widetilde{W}=t_0\ \tr \left( -{1\over 3} \gamma^3 Y^3 + t_1^2 
\gamma Y \right)-h\mu^2 \  \tr \ M -h'\mu'^2\ \tr M'X_2.
}
The new scale $\Lambda_{\rm susy}$ for the $U(2N_2-N_1)$ gauge group
after decoupling $(q_{12}, q_{21})$  
is given by the matching condition at the mass
scale $h M$ as
\eqn\magneticsusyvac{
 (hM)^{N_2}
\ \widetilde{\Lambda}_1^{3N_2-2N_1} =
\Lambda_{\rm susy}^{2(2N_2-N_1)}.
}
On the other hand, the coupling constant $g_2$ for $U(N_2)$ stops running
at the energy $hM$ since those charged with respect to this gauge
group are $M, M'$ and $X_3$ and they make the same matter content
as that for the ${\cal N}=4$ supersymmetric gauge theory albeit with
the different superpotential. The coupling constant $g_2$ 
is small since we assume $h M \ll \widetilde{\Lambda}_2$
so that we stay well below the Landau pole. Thus, we can 
integrate out $M'$ and $X_2$ as they are free in the limit of small 
$g_2$. 

The vacuum expectation value \ydiag\ of $Y$ breaks the gauge symmetry as
$$ U(2N_2-N_1)\rightarrow U(r_1)\times U(r_2)$$ 
with $r_1+r_2=2N_2-N_1$. The strong coupling scales 
$\Lambda_{U(r_a)}$ for $U(r_a)$ ($a=1,2$) are determined  by
the matching condition at the threshold as
$${\eqalign{
\Lambda_{U(r_1)}^3 
&=\gamma^3 t_0 \left({t_1\over \gamma}\right)^{r_1-2r_2 \over r_1}
\Lambda_{\rm susy}^{2(2N_2-N_1)\over r_1}
=\gamma^3 t_0\left({t_1\over \gamma}\right)^{r_1-2r_2 \over r_1}
 \widetilde{\Lambda}_1^{3N_2 -2N_1\over r_1}(h M)^{N_2 \over r_1} ,\cr
\Lambda_{U(r_2)}^3 
&=\gamma^3 t_0 \left({t_1\over \gamma}\right)^{r_2-2r_1 \over r_2}
\Lambda_{\rm susy}^{2(2N_2-N_1)\over r_2}
=\gamma^3 t_0\left({t_1\over \gamma}\right)^{r_2-2r_1 \over r_2}
\widetilde{\Lambda}_1^{3N_2 -2N_1\over r_2} (h M)^{N_2 \over r_2} .
}
}$$
The effective superpotential $W_{{\rm eff}}$ for $M$ would then be 
a function of these scales of the form: 
$$ W_{\rm eff} = - h \mu^2 M + W_{\rm nonpert}(\Lambda_{U(r_1)}, 
\Lambda_{U(r_2)}). $$
For fixed $t_0, t_1$ and $\widetilde{\Lambda}_1$, the second term
in the above is a non-trivial function of $M$. If it is sufficiently
generic, there are solutions to the $F$-term condition,
\eqn\canbesolved{  {\partial W_{\rm eff} \over \partial M}  = - h \mu^2
 + {\partial W_{\rm nonpert}\over \partial M} = 0.}
Thus, the non-perturbative $U(2N_2-N_1)$ gauge dynamics can 
restore the supersymmetry for large $M$. 

To make quantitative estimate of $W_{{\rm nonpert}}$, let us assume 
that
\eqn\smalllambda{ \Lambda_{\rm susy} \ll {t_1 \over \gamma}.}
In this limit, fields that are bifundamental in 
$U(r_1) \otimes U(r_2)$ become heavier than the 
gauge theory scale and decouple. Thus, we can treat the
two gauge group factors separately and estimate the effective superpotential
as 
\eqn\npeff{
W_{\rm nonpert}\simeq \sum_{a=1,2} r_a \Lambda_{U(r_a)}^{3}+{\cal O} 
\left( \Lambda_{U(r_a)}^6 {\gamma^3 \over t_1^3}\right).
}
Without  loss of generality, we can assume $r_1 \leq r_2$. 
If $r_1< r_2$,  we have
$${
{\Lambda_{U(r_1)}^3\over \Lambda_{U(r_2)}^3}
=\left({t_1 \over \gamma \Lambda_{\rm susy}} 
\right)^{2(r_1^2-r_2^2)\over r_1 r_2} \ll 1,
}$$
where we used \smalllambda . In this case, we can ignore $\Lambda_{U(r_1)}$ 
in the effective superpotential \npeff . On
the other hand, $\Lambda_{U(r_1)}
=\Lambda_{U(r_2)}$ for $r_1=r_2$. Thus, in either case, 
$$\eqalign{
W_{\rm nonpert} \sim  r_2 \Lambda_{U(r_2)}^3 
&=r_2 \gamma^3 t_0 \left({t_1\over \gamma}\right)^{r_2-2r_1 \over r_2}
\Lambda_{\rm susy}^{2(2N_2-N_1)\over r_2} \cr
& =r_2 \gamma^3 t_0\left({t_1\over \gamma}\right)^{r_2-2r_1 \over r_2}
 \widetilde{\Lambda}_1^{3N_2 -2N_1\over r_2} (h M)^{N_2 \over r_2}.
}$$
With this non-perturbative term, the supersymmetry condition
indeed has solutions of the form 
$${
h  M =  \zeta^{{2r_2\over N_2-r_2}}\eta^{{2r_1-r_2 \over N_2-r_2}} \
\widetilde{\Lambda}_1 \ \unit_{N_2},}$$
where
$$ \zeta^2={1 \over N_2 \gamma^3 t_0 } 
\left({\mu \over \widetilde{\Lambda}_1}\right)^2,
\ \ \eta= {t_1 \over\gamma \widetilde{\Lambda}_1}.
$$

Let us discuss the region of validity of our estimate. In 
order for our analysis to be consistent, 
we require the relevant particle masses, $h M$ and $t_1/\gamma$,
are below the Landau pole at $\widetilde{\Lambda}_1$.
This means $\eta \ll 1$ and 
\eqn\zetaetatwo{ \zeta^{2r_2} \eta^{2r_1-r_2} \ll 1.}
If we require that this is satisfied for all the gauge symmetry breaking
pattern, the strongest constraint
from \zetaetatwo\ comes from the case when $r_1=0$, $r_2=2N_2-N_1$
(to be precise, this is allowed when $N_2$ is even). Thus, 
we have,
$$\zeta \ll \eta^{{1\over 2}} \ll 1.$$
In addition, we require \smalllambda\ so that we can use
the explicit expression \npeff\ for the non-perturbative
effective potential. This condition can be expressed in 
terms of $\zeta$ and $\eta$ as,
\eqn\zetaetaone{ \zeta \ll \eta^{{N_1\over N_2} -{1\over 2}}.}
Since $N_1 > N_2$, this condition is stronger than $\zeta \ll 
\eta^{1\over 2}$. Thus, all the inequalities are satisfied
when 
\eqn\zetaetathree{ \zeta \ll \eta^{{N_1\over  N_2} -{1\over 2}} \ll 1.}  

We can now estimate the decay rate of each meta-stable vacuum following \ISS. 
At the semi-classical level, the decay probability is proportional to 
$e^{-S}$ where $S$ is the Euclidean action for the decay process. 
Using the formula in \triangle, we find 
\eqn\bouce{\eqalign{
S \sim {(\Delta \Phi)^4\over V_+}& 
= {(\tr h M)^4 \over (N_1-N_2)h^2 \mu^4} \cr 
&={N_2^2 \over (N_1-N_2)h^2 t_0^2 \gamma^6} 
\left({\eta^{2r_1-r_2}\over \zeta^{N_2-3r_2} }  
\right)^{4\over N_2-r_2}.
}}
where $\Delta \Phi$ is the order of the difference of the vacuum expectations
values at the meta-stable vacua and at the supersymmetric vacua, 
and $V_+=(N_1-N_2) |h^2 \mu^4|$. The meta-stable vacua are long-lived if $S \gg 1$,
which means
\eqn\lifetimeconstII{
\eta^{2r_1-r_2} \gg \zeta^{N_2-3r_2}.
}
Requiring this for all $r_1 \leq r_2$ with $r_1+ r_2 = 2N_1-N_2$
and combining  it with \zetaetathree , we find,
$$ \eqalign{
\eta^{{2N_2-N_2 \over 5N_2-3N_1}} \ll &\zeta \ll
\eta^{{N_1 \over N_2} -{1\over 2}}\ll 1~~~~~({\rm if}~N_2 > {3\over 5}N_1) \cr 
&\zeta \ll
\eta^{{N_1 \over N_2} -{1\over 2}}\ll 1~~~~~({\rm if}~N_2 \leq {3\over 5}N_1). }
$$
These conditions also allow us to ignore higher order 
correction to the K\"ahler potential even though the 
non-perturbative corrections to the superpotential are included.

One may be concerned about transitions among supersymmetry
breaking vacua. Since they all have the same energy, their
transition probabilities are all zero. 

\newsec{$SU(N_1)\times SU(N_2)$ Gauge Theory}

Let us briefly describe our result for the quiver gauge theory
with the gauge group $SU(N_1)\times SU(N_2)$. 
In this case,  we can add a quadratic term to $W_1(X_1)$ as 
an independent superpotential term.  
It cannot be removed by constant shift of $X_1$ and $X_2$ 
because of the tracelessness condition on them. Let us parametrize
$W_1(X_1)$ as
$$ W_1(X_1) = t_0 \ \tr \left( {1\over 3} X_1^3
+ {t_2\over 2} X_1^2 \right), $$
with the condition $\tr X_1 = 0$. 
Assuming $\Lambda_2 \ll \Lambda_1$, we can use the 
magnetic dual with respect to $SU(N_1)$ identified in \KSS . 
Repeating the analysis in section 2.2 and using the
same notation, we find
that the tree-level vacua in the dual description 
break the gauge symmetry to $SU(r_1) \times SU(r_2)
\times SU(N_1-N_2) \times U(1)^2$ with $r_1 +r_2 = 2N_2-N_1$ by
the vacuum expectation value of $Y$
as
\eqn\newdiag{ \gamma Y = {\rm diag}( \lambda, \cdots, \lambda, -\lambda,
\dots, - \lambda),}
where 
\eqn\whatlambda{ \lambda = {N_1 t_2 \over 2 (r_2 - r_1)}, }
and $r_2$ is the number of minus signs in \newdiag .
Without a loss of generality, we assume $r_1 < r_2$. 
Note that the eigenvalues of $\gamma Y$ depend on $(r_1, r_2)$,
namely on the choice of vacuum. This is in contract to
the $U(N_1) \times U(N_2)$, where the eigenvalues
are given by $\pm t_1$. 

Another difference is that there is an extra flat direction 
in addition to the $\Phi_0$ as in \Vmin. 
By using the complexified gauge group $SL(2N_2-N_1, C)$,
we can diagonalize $\varphi_0$ and $\widetilde{\varphi}_0$
obeying $\widetilde{\varphi}_0 \varphi_0 = \mu^2 \ \unit_{2N_2-N_1}$. 
However, unlike the previous case,
we cannot set $\varphi_0 = \widetilde{\varphi}_0$ since
$SL(2N_2-N_1, C)$ does not contain the overall scaling.
The best we can do with this gauge group is
$${
\varphi_0=\mu e^\theta \ \unit, ~~\widetilde{\varphi}_0 = 
\mu e^{-\theta} \ \unit. 
}$$ 
This $\theta$ is the additional flat direction for the tree-level
potential.

The one-loop effective potential for $\Phi_0$ and $\theta$
can be evaluated in the same way as in the $U(N_1) \times U(N_2)$
model, and we have checked that both $\Phi_0$ and $\theta$
are stabilized. Thus, all the flat directions
are lifted at one-loop. A new feature of this model is that
the one-loop effective action depends on the choice of vacuum.
This follows from the fact that the vacuum expectation 
value of $\gamma Y$ depends on $(r_2 -r_1)$ as in \whatlambda .
In particular, the meta-stable vacua have different one-loop
vacuum energies.

We have also estimated the lifetimes of these supersymmetry breaking
meta-stable vacua. 
The analysis is the same as $U(N_1)\times U(N_2)$ case with 
the substitution $t_1 \rightarrow \lambda$ since
the $U(1)$ factors in $U(N_1)$ does not affect the 
non-perturbative superpotential. Thus we conclude that the 
decay rates of the meta-stable vacua into the supersymmetric
vacua can be made parametrically small in this model also. 
In this case, there may be non-zero transition probabilities
between supersymmetry breaking vacua. 

\bigskip
\bigskip
\centerline{\bf Acknowledgments}
\bigskip 
 
This research is supported in part by 
DOE grant DE-FG03-92-ER40701. 
Y.O. is also supported in part by the JSPS Fellowship for Research Abroad. 
We would like to thank Tom Banks, Michael Dine, Michael Graesser, 
Ken Intriligator, Yu Nakayama, Masaki Shigemori, David Shih, 
Yuji Tachikawa, Taizan Watari, and Futoshi Yagi
for discussions.


\listrefs

\end